\documentclass[aps,preprintnumbers,superscriptaddress,twocolumn]{revtex4}
\usepackage{amsfonts,amssymb,amsmath}            
\usepackage[pdftex]{graphics,graphicx,epsfig}            
\usepackage{amsthm}
\usepackage{color}
\def\identity{\leavevmode\hbox{\small1\kern-3.8pt\normalsize1}}
\newcommand{\beq}{\begin{equation}}
\newcommand{\eeq}{\end{equation}}
\newcommand{\beqa}{\begin{eqnarray}}
\newcommand{\eeqa}{\end{eqnarray}}

\newcommand{\ket}[1]{|#1\rangle}
\newcommand{\bra}[1]{\langle#1|}

\newcommand{\half}{\mbox{$\textstyle \frac{1}{2}$}}
\newcommand{\proj}[1]{\ket{#1}\bra{#1}}
\newcommand{\Tr}{\text{Tr}}

\begin{document}
\title{A Witness of Multipartite Entanglement Strata}

\author{Dagomir Kaszlikowski}
\affiliation{Centre for Quantum Technologies, National University of
Singapore, 3 Science Drive 2, Singapore 117543}
\author{Alastair Kay}
\affiliation{Centre for Quantum Computation, Department of Applied
Mathematics and Theoretical Physics, University of Cambridge, Cambridge CB3
0WA, UK}
\affiliation{Centre for Quantum Technologies, National University of
Singapore, 3 Science Drive 2, Singapore 117543}
\date{\today}

\begin{abstract}
We describe an entanglement witness for $N$-qubit mixed states based on 
the properties of $N$-point correlation functions. Depending on the degree of violation, this witness can guarantee that no more than $M$ qubits are separable from the rest of the state for any $M\leq N$, or that there is some genuine $M$-party or greater multipartite entanglement present. We illustrate the use our criterion by investigating the existence of entanglement in thermal stabilizer states, where we demonstrate that the witness is capable of witnessing bound-entangled states. Intriguingly, this entanglement can be shown to persist in the thermodynamic limit at arbitrary temperature.
\end{abstract}
\maketitle

\section{Introduction}

Recently, there has been growing interest in the entanglement of large many-body systems. The reasons for this are many and varied. From the quantum information perspective, quantum computation intrinsically involves the use of large arrays of entangled qubits, including the one-way model of computation \cite{cluster} where the initial, cluster state, of the system is considered a multipartite entangled resource for computation. This model for computation has the potential to enable quantum information processing in many physical systems which were otherwise considered infeasible. Generating the cluster state and finding signatures with which to recognise its successful generation along with protecting and recovering it from the effects of noise are thus matters of great importance.
Of more immediate relevance to condensed matter theory and experiments is the study of, for example, how thermodynamic parameters of many-body systems as well as phase transitions can be directly related to entanglement \cite{thermal, phase}. However, it is still not clear if this relation always holds and if it is of any real significance to our understanding of the physics of large quantum systems. 

To be able to better investigate such properties both theoretically and experimentally, it is interesting to develop techniques for entanglement detection in many-qubit systems. The most straightforward approach is via the so-called entanglement witnesses. Entanglement witnesses are Hermitian operators such that their mean value with an arbitrary separable state is always less than one. Thus, if the observed value exceeds one, we are certain that a given state is entangled.  The significance of the entanglement witness approach is that, in principle, it can be experimentally implemented and that one can find entanglement witnesses tailored to detect multi-partite entanglement \cite{toth,brandao, toth2} as well as entanglement strong enough to violate some Bell inequalities. It was shown in Ref. \cite{games} that one can perform certain computational tasks more efficiently than using classical resources only if entanglement violating Bell inequalities is available as a resource.

There is little literature dealing with entanglement witnesses for many qubit systems capable of detecting genuine multipartite entanglement \cite{lit1,lit2,lit3,toth2}. The usual approach (see, for instance, Ref. \cite{toth2}) is to find an entanglement witness tailored for a specific state and then show that it can be measured locally. Usually such witnesses are capable of detecting entanglement only in the neighbourhood of the state for which they have been optimized. In this
paper we derive a witness that is capable of detecting multipartite entanglement for a wide set of multi qubit states. Moreover, the number of different settings required to measure our witness does not have to increase exponentially with the
number of qubits (although generally it does), which is often the case \cite{lit1,lit2,lit3}, with the notable exceptions of Refs. \cite{toth2, vlatko}.

In this paper, we find a family of entanglement witnesses for $N$ qubits, the mean value of which solely depends on $N$-point correlation functions. We derive a relatively simple lower bound for the witness and demonstrate its calculation with the examples of thermal stabilizer states and the thermal single-excitation Bose-Hubbard model to obtain a range of temperatures for which entanglement exists. We also relate the lower bound to the violation of Bell inequalities from the so-called Werner-Wolf, Zukowski-Bruckner (WWZB) family \cite{WWZB}. We show that depending on the different level (strata) of violation of the inequalities, we can assign lower bounds to the degree of entanglement present.

\section{Witnessing Separability and Multipartite Entanglement}

In \cite{valerio}, a convenient parametrization of all two-setting Bell inequalities of the WWZB type was developed by considering the Hermitian operator
\begin{equation}
W = \half\sum_{\vec{k}\in\{0,1\}^N} b_{\vec{k}}\left (Q^+_{\vec{k}}-Q^-_{\vec{k}}\right ), \label{eqn:witness}
\end{equation}
subject to a variety of constraints on the coefficients $b_{\vec{k}}$. The operators  $Q^{\pm}_{\vec{k}}=\proj{G^{\pm}_{\vec{k}}}$ are orthogonal projectors on the generalized GHZ states $\ket{G^{\pm}_{\vec{k}}} = \frac{1}{\sqrt{2}}(|\vec{k}\rangle \pm \sigma_x^{\otimes N}|\vec{k}\rangle)$. For convenience, we can express $ \sigma_x^{\otimes N}|\vec{k}\rangle = |\vec{k}'\rangle$. We will now investigate the properties of $W$ devoid from the restrictions imposed in \cite{valerio}, except that we shall require $W$ to be an entanglement witness.

\subsection{Conditions for Full Separability}

The family of operators $\hat{U}W\hat{U}^{\dagger}$, where $\hat{U} = \prod_{n=1}^N U^{(n)}$ and $U^{(l)}$ is an arbitrary $SU(2)$ transformation on qubit $l$, becomes a family of entanglement witnesses if 
\begin{equation} 
\sum_{\vec{k}} |b_{\vec{k}}|\leq 2^N,
\label{wit}
\end{equation}
since, if this condition holds, one has 
\begin{equation}
\langle\psi_{sep}| \hat{U} W\hat{U}^{\dagger}|\psi_{sep}\rangle \leq 1	\label{eqn:sep}
\end{equation}
for an arbitrary pure and fully separable state 
\begin{eqnarray}
&&|\psi_{sep}\rangle = \bigotimes_{j=1}^N V^{(j)}|\vec{0}\rangle.	\nonumber
\end{eqnarray}
In the above formula, $V^{(j)}$ is an $SU(2)$ transformation which can be parametrized by 
$
V^{(j)} = e^{-i\theta\vec n_j\cdot\vec \sigma},
$
where ${\vec n}$ is a unit vector.
To prove Eqn.~(\ref{eqn:sep}), firstly observe that the unitaries $U^{(l)}$ can be absorbed in $V^{(l)}$, so that we only have to show that
\begin{eqnarray}
&&\max_{|\psi_{sep}\rangle}\langle\psi_{sep}|W|\psi_{sep}\rangle =\nonumber\\
&& \max_{|\psi_{sep}\rangle}\sum_{\vec{k}} b_{\vec{k}} \text{Re}{\left(\langle\psi_{sep}|\vec{k}\rangle\langle\vec{k}'|\psi_{sep}\rangle\right)}\leq 1,
\label{re}
\end{eqnarray}
where we've made use of the expansion $W=\frac{1}{2}\sum_{\vec{k}}b_{\vec{k}}\left(\ket{\vec{k}}\bra{\vec{k}'}+\ket{\vec{k}'}\bra{\vec{k}}\right)$.
The properties of $V^{(j)}$ impose that  
\begin{eqnarray}
\text{Re}\left(\langle\psi_{sep}|\vec{0}\rangle\langle\vec{0}'|\psi_{sep}\rangle\right) &=&
\text{Re}\left(\prod_{i=1}^N\bra{0}V^{(i)\dagger}\ket{0}\bra{1}V^{(i)}\ket{0}\right),\nonumber
\end{eqnarray}
where we have also absorbed the $\sigma_x$ rotations from $\vec{k}$ to $\vec{0}$ into the $V^{(j)}$.
Therefore, the maximum over $\vec n$ and $\theta$ of Eqn.~(\ref{re}) yields 
\begin{equation} 
\max_{|\psi_{sep}\rangle}\langle\psi_{sep}|W|\psi_{sep}\rangle = \frac{1}{2^N}\sum_{\vec{k}}|b_{\vec{k}}|,
\end{equation}
with values $n_z=n_x=\half$ and $\theta=0$. This is not greater than 1 when Eqn.~(\ref{wit}) holds. Proving this for pure states is sufficient since the convexity of the mixed separable states implies that the optimum will be given by a pure state.

\subsection{Entanglement Witnesses}

In order to detect entanglement, we have to calculate
$$
\Tr(W\rho)=\max_{\hat U}\sum_{\vec k}b_{\vec k}\text{Re}(\bra{\vec k}\hat U\rho\hat U^\dagger\ket{\vec k'})
$$
subject to the constraint (\ref{wit}). We shall now consider two specific cases of entanglement witness $W$, by selecting two specific sets of $b_{\vec k}$. The first, $W_A$, is the strongest witness of this class, and is selected by finding the value of ${\vec k}_0$ which maximizes $\max_{\hat U}|\text{Re}(\bra{\vec k}\hat U\rho\hat U^\dagger\ket{\vec k'})|$, setting $b_{\vec{k}_0}$ to $2^N$, and all others to 0.

In general, it is likely that the maximization over ${\vec k}$ will be difficult to do, so we choose another witness, $W_B$, which will give a lower bound to this value. We select
\begin{equation}
b_{\vec{k}} = 2^N\lambda_{\vec{k}} \left(\sum_{\vec{l}} \lambda_{\vec{l}}\right)^{-1},	\label{eqn:W_b}
\end{equation}
where
$$
\lambda_{\vec{k}} = \sum_{\vec{l}}(-1)^{\vec{k}\cdot\vec{l}}\cos{\left(\frac{\pi}{2}|\vec{l}| \right)}T_{\vec{l}}.
$$
The number $T_{\vec{l}}= \Tr(\hat{U}\sigma_{\vec{l}}\hat{U}^{\dagger}\rho)$ is the average value of spin measurements along the directions given by $O_{n}\hat{x},O_n\hat{y}$ ($n=1,\dots,N$), where $O_n$ is an orthogonal representation of $U_n$.  We see that the only relevant $N$-point correlation functions are those for which $|\vec{l}|$, i.e., the number of $\sigma_y$ in $\sigma_{\vec{l}}$ before the local unitary operation $\hat{U}$, is an even number. The choice of $b_{\vec k}$ is clearly sub-optimal, so $\Tr(W_A\rho)\geq\Tr(W_B\rho)$. The reason for this particular choice is that
\begin{equation}
\Tr(W_B\rho)=\max_{\hat{U}}\left(\sum_{\vec{l}\in even}T^2_{\vec{l}}\right),	\label{eqn:W_b2}
\end{equation}
as will be proved in the Appendix.

Our starting point was motivated by the fact that it was shown in Ref.~\cite{valerio} that the family of operators $\hat{U}W\hat{U}^{\dagger}$ subject to the constraint $\sum_{\vec{k}}b_{\vec{k}}^2=2^N$ and some complicated additional constraints on the signs of the $b_{\vec{k}}$'s coincides with the family of all two-setting Bell inequalities of the WWZB type \cite{WWZB}. Therefore, if for some state $\rho$ one maximizes $\Tr(W_{\hat{U}}\rho)$ only under the constraint $\sum_{\vec{k}}b_{\vec{k}}^2=2^N$, one gets the upper bound for violation of two-setting WWZB inequalities. If this maximum is larger than one, nothing conclusive can be said about violation of two-setting WWZB inequalities for the state $\rho$. However, if $\Tr(W_{\hat{U}}\rho)\leq 1$, one concludes that the state $\rho$ cannot violate any of the two-setting inequalities from the WWZB family.

Specifically, selecting $b_k=2^N\lambda_k/\sqrt{\sum_k\lambda_k^2}$ yields the maximum under the constraint $\sum_{\vec{k}}b_{\vec{k}}^2=2^N$, and gives
\begin{eqnarray}
W_C=\sqrt{W_B}=\max_{\hat{U}}{ \sqrt{ \sum_{\vec{l}\in even}T_{\vec{l}}^2}}.
\end{eqnarray}
Thus, if $W_C\leq 1$, which is equivalent to the condition $W_B\leq 1$, one cannot violate two-setting Bell inequalities from the WWZB family. This is in agreement with the necessary condition given in \cite{WWZB} for not violating two-setting WWZB inequalities.  Interestingly, $W_B>1$ is a sufficient condition for violation of the multi-setting WWZB Bell inequalities presented in Ref. \cite{ZB}. 

\subsection{$W_A$ as a witness of partial separability}

Let us examine the witness $W_A$ more carefully, detailing its effect on pure states $\rho=\proj{\psi}$.
\begin{eqnarray}
\Tr(W_A\rho)&=&2^N\max_{k,\hat U}\text{Re}(\bra{k}\hat U\proj{\psi}\hat U^\dagger\ket{k'})	\nonumber\\
&=&2^N\max_{\hat U}\text{Re}\left(\bra{0}^{\otimes N}\hat U\proj{\psi}\hat U^\dagger\ket{1}^{\otimes N}\right).	\nonumber
\end{eqnarray}
We can now split $\ket{\psi}$ into a part $\ket{\tilde \psi}$ acting on $M$ qubits, and $\ket{\tilde \psi_{sep}}$, a fully separable state on the other $N-M$ qubits. The witness does not change under permutations, so without loss of generality, we can take the qubits of $\ket{\tilde \psi}$ to be the first $M$ qubits. Having already established that
$$
\max_{\hat U}\text{Re}(\bra{0}^{\otimes N-M}\hat U\proj{\tilde\psi_{sep}}\hat U^\dagger\ket{1}^{\otimes N-M})=\frac{1}{2^{N-M}},
$$
we see that the maximum value of $\Tr(W_A\rho)$ where no more than $M$ qubits are entangled is given by $\Tr(W_A\proj{\tilde \psi})$, our entanglement witness on $M$ qubits. Hence, the largest possible value is $2^{M-1}$, where $\ket{\tilde \psi}=(\ket{0}^{\otimes M}+\ket{1}^{\otimes M})/\sqrt{2}$. We conclude that if $\Tr(W_A\rho)>2^{M-1}$, fewer than $N-M$ qubits are separable. Again, by convexity, the results also apply to mixed states, even though we only performed the calculation for pure states. So, our entanglement witness witnesses not only the fact that there is some entanglement, but witnesses that at least $M+1$ qubits are entangled, although this currently makes no statement about the type of entanglement present.

\subsection{$W_A$ as a witness of multipartite entanglement}

Given that we can witness the fact that many qubits are entangled, it would also be interesting if we can witness different types of multipartite entanglement. The first step in this process is to determine if $M$-party (or greater) multipartite entanglement is involved. Again, we consider $W_A$ acting on pure states. Since we know that GHZ states maximize the value of the witness (by design), then if $M$-partite entanglement is involved (assuming $N/M$ is an integer), the maximum violation of $W_A$ must be given by
$
\ket{\psi}=\ket{\tilde \psi}^{\otimes N/M},
$ 
where $\ket{\tilde \psi}=(\ket{0}^{\otimes M}+\ket{1}^{\otimes M})/\sqrt{2}$. We find that
$$
\Tr(W_A\rho)=\Tr(W_A\proj{\tilde \psi})^{N/M}=2^{N-N/M}.
$$
Thus, if $\Tr(W_A\rho)>2^{N-N/M}$, the multipartite entanglement that is involved must be at least $(M+1)$-partite. If $N/M$ is not an integer, the maximal violation is given by $\lfloor N/M\rfloor$ copies of an $M$-qubit GHZ state, and a single $M(N/M-\lfloor N/M\rfloor)$-qubit GHZ state, giving a violation of
$$
\Tr(W_A\rho)=2^{N-1-\lfloor N/M\rfloor}.
$$
Note that for large $M$, several values give the same threshold. For example, all values $N/2<M\leq N$ give a violation of $2^{N-2}$. It is also worth observing that since $W_B$ provides a lower bound to $W_A$, $W_B$ can be assigned the same interpretation for violating the strata of thresholds.

Finally, one can also take a more specialized approach to using the witness, developing specific strategies to resolve different types of entanglement. For example, were one to be presented with an $N$-qubit pure state $\ket{\psi}$, and promised that it is an $M$-qubit $W$-state, with all other qubits separable, then we might like to determine the value of $M$. This can be achieved by measuring the value of $W_A$ on $N-2$ different partitions. By measuring $W_A$ on a subset of qubits, if that subset entirely encompasses the $W$-state, we get value $\Tr(W_A\rho)=2$. However, if it only encompasses $R$ of the $M$ qubits, the value is $1+R/M$. Thus a systematic search using subsets of qubits $1$ to $n$ for $2<n\leq N$, the changes in value can be detected, and $M$ determined.

\subsection{Summary}

In this section, we have presented an entanglement witness, $W_A$. Depending on the degree of violation, it detects not only full separability, but can give an upper bound on the number of separable qubits. The extent of the violation also serves to witness the presence of genuine multipartite entanglement of differing types, tuned most specifically to be sensitive to GHZ-like entanglement. Since the optimization involved in calculating $W_A$ is typically difficult, we presented a sub-optimal witness $W_B$ that also possesses these properties, and is easier to calculate.
Furthermore, the witness provides an upper bound to the violation of WWZB inequalities -- if $\Tr(W_B\rho)\leq 1$ is not violated, then no WWZB inequality can be violated. 

\begin{figure}
\begin{center}
\includegraphics[width=0.45\textwidth]{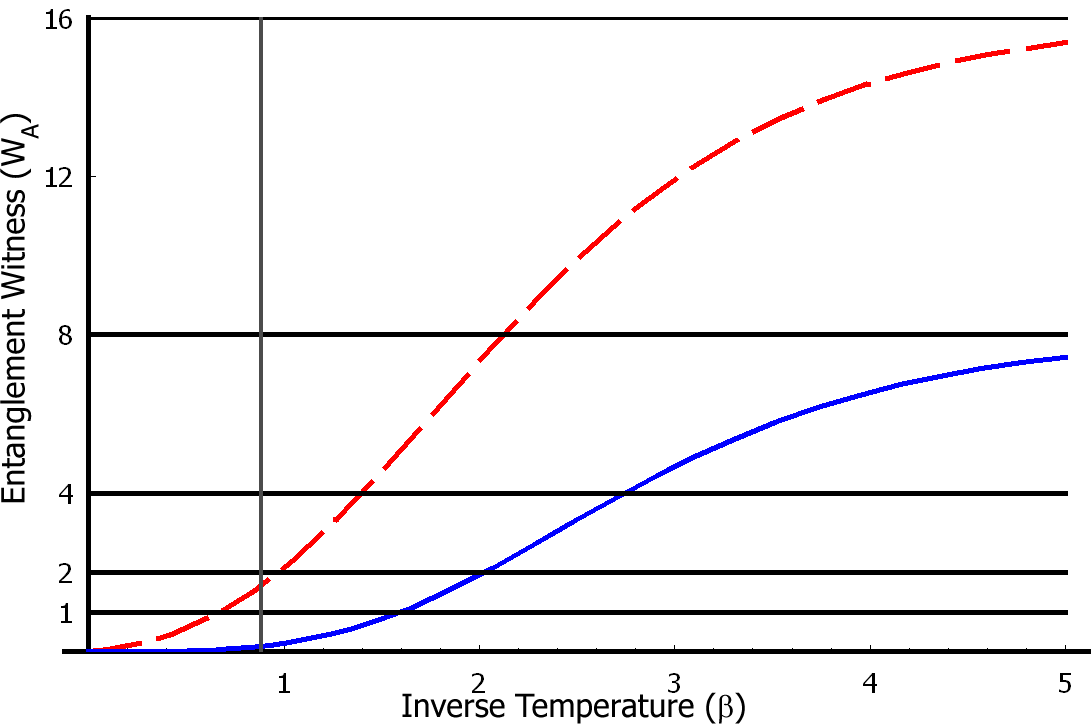}
\end{center}
\caption{A plot of the violation of our entanglement witness as a function of inverse temperature, $\beta$, for the 5-qubit GHZ state (dashed) and 7-qubit cluster state (solid). Indicated in black are the witnesses as to the minimum number of qubits, $\{2,3,4,5\}$, that are entangled for values $\{1,2,4,8\}$ respectively. The values of $\{1,4,8\}$ also witness, for the GHZ state, the existence of $\{2,3,5\}$-partite entanglement. The vertical grey line indicates the temperature below which which either state can be purified by a genuine multipartite protocol.} \label{fig:main}
\end{figure}

In the following section, we will apply the witness to some simple examples. This will enable us to demonstrate some of the properties of the entanglement witness. For example, it is capable of witnessing bound entanglement. We will also be able to compare the two witnesses and see how tight a lower bound is provided by $W_B$. A summary of the results is depicted in Fig.~\ref{fig:main}.

\section{Examples}

\subsection{Thermal Stabilizer States}

In this section of the paper we analyze entanglement in thermal mixtures of the stabilizer Hamiltonians with the 
help of the entanglement witnesses $W_A$ and $W_B$. The stabilizer states, which are the eigenstates of an associated Hamiltonian $H=-\half\sum_{n=1}^{N}K_n$, where $[K_n,K_m]=0$ and $\Tr(K_n)=0$, enable a particularly simple description of the thermal state. They are of particular interest because special cases of the stabilizer states include many of the important states in quantum information such as GHZ states, cluster states and error correcting codes. The examples that we present serve to illustrate the variety of properties that can be detected by our entanglement witness. 

The thermal state of $H$ can be expanded due to the commutation of the operators, and the fact that $K_n^2=\identity$, such that
\begin{equation}
\rho=\frac{e^{-\beta H}}{\Tr(e^{-\beta H})}=\frac{1}{2^N}\prod_{n=1}^N(\identity+\tanh(\beta/2)K_n).
\end{equation}
where, as usual, $\beta^{-1} =k_BT$, $k_B$ is the Boltzmann constant and $T$ the temperature of the system \footnote{We are assuming that the ground state is uniquely defined i.e.~there is no degeneracy}.
To evaluate the witnesses, we need to find the values of the $N$-point correlations functions and optimize over all possible local bases, and to achieve this we simply need to consider the products of the operators $K_n$.

While we have been considering calculating $W_B$ as a lower bound for $W_A$ because, in general, $W_A$ is more difficult to calculate, in the following examples, it turns out to be no harder than the calculation of $W_B$. Let's assume that we have a minimal sequence of products of stabilizers that gives an $N$-body correlator, and perform local rotations such that this is the correlator of all $\sigma_x$s. Additionally, assume that all other $N$-body correlators will come from multiplying additional terms with this one.
The local rotations also ensure that these terms that we multiply by are $\sigma_z$s, such that the outcome is  $i\sigma_y$. Thus, the overall phase is given by $i^{|\vec l|}=\cos{\left(\frac{\pi}{2}|\vec{l}|\right)}$ where $\vec l$ denotes the positions of the $\sigma_z$s. This $\cos{\left(\frac{\pi}{2}|\vec{l}|\right)}$ in $T_{\vec l}$ multiplies the identical term in the expression for $\lambda_{\vec k}$, and gives $+1$ since $|\vec{l}|$ is even. Hence, all the terms are positive quantities, except for the term $(-1)^{\vec{l}\cdot \vec{k}}$, which can be set positive for all $\vec{l}$ by choosing $\vec k=0$. Consequently, for all these cases, we can achieve
$$
\Tr(W\rho)=\sum_{\vec{k}}|T_{\vec{k}}|
$$
which is also an upper bound to the value of $W_A$, and is hence the optimal choice. This presents the opportunity of assessing how well $W_B$ performs as a lower bound to $W_A$. In fact, given the assumption that all correlators are the result of the products of stabilizers, then all $T_{\vec l}$ are powers of $\tanh(\beta/2)$, and hence if $W_A$ determines a critical temperature of $\beta_A$, then
$$
\tanh(\beta_A/2)=\tanh^2(\beta_B/2).
$$
At the extreme of large $\beta_A$, $\beta_B=\beta_A\ln(2)$, and for small $\beta_A$, $\beta_B=\sqrt{2\beta_A}$, so the bound does not seem wholly unreasonable.

\subsection{GHZ Hamiltonian}

By choosing the following stabilizers,
\begin{eqnarray}
K_1&=&\prod_{n=1}^N\sigma_x^{(n)}	\nonumber\\
K_n&=&\sigma_x^{(1)}\sigma^{(n)}_z,
\end{eqnarray}
the ground state of $H$ is the $N$-qubit GHZ state $(\ket{0}^{\otimes N}+\ket{1}^{\otimes N})/\sqrt{2}$. The excited states constitute the ground state with local operators $\sigma_z^{(1)}$ or $\sigma_x^{(n\neq 1)}$ applied. Since our entanglement witness is formed from projectors on GHZ states, this is a natural test candidate.

It is evident that the optimal choice of basis consists of the eigenstates of the Pauli operators $\sigma_x$ and $\sigma_y$. This is particularly clear in the case of odd $N$, where the $N$-body correlator necessarily includes a term due to $K_1$ ($\prod_{n\neq 1}K_n$ creates a correlator on all other qubits, but not qubit 1). Once $K_1$ is included, multiplying by any other stabilizer necessarily gives an $N$-body term formed of $\sigma_x$ and $\sigma_y$ terms. All such products introduce an even number of $\sigma_y$s so that we get
\begin{eqnarray}
Tr(W_A\rho)&=&\sum_{n=0}^{N-1}\binom{N-1}{n}\tanh^{n+1}(\beta/2)\nonumber\\
&=&\tanh(\beta/2)\left(1+\tanh(\beta/2)\right)^{N-1}\nonumber\\
Tr(W_B\rho)&=&\tanh^2(\beta/2)\left(1+\tanh^2(\beta/2)\right)^{N-1}
\end{eqnarray}
In the thermodynamic limit, i.e., $N\rightarrow \infty$, we find the limit of $Tr(W_A\rho)=1$ at $\tanh(\beta/2)\approx 1/\sqrt{N}$. Since this shows that $\beta\rightarrow 0$, it has the interpretation that large systems of this form are always entangled. However, we know from \cite{AK1} that distillation is impossible if $\beta\leq \ln\left(\sqrt{2}+1\right)$, which causes us to conclude that this persistent entanglement is bound entanglement i.e.~entanglement that cannot be distilled by a multipartite distillation protocol. This proof arises from considering a particular bipartition of the qubits, and showing that distillation across that bipartition is impossible. However, there exist other bipartitions across which distillation is possible. Witnesses, including Bell tests, for bound entanglement have previously been demonstrated \cite{wolfgang, tony, dag}. What is perhaps most remarkable about this system is that entanglement becomes more persistent in larger systems. Unfortunately, the $N$-body term $K_1$ in the Hamiltonian is not particularly physical, so we should not necessarily expect to see the consequences in real-world systems involving local interactions.

\subsection{Cluster State}

Another example of a stabilizer state is the cluster state. We shall restrict to the one-dimensional version, where the stabilizers are defined as
\begin{eqnarray}
K_1&=&\sigma_x^{(1)}\sigma_z^{(2)}	\nonumber\\
K_N&=&\sigma^{(N-1)}_z\sigma^{(N)}_x	\nonumber\\
K_n&=&\sigma_z^{(n-1)}\sigma_x^{(n)}\sigma^{(n+1)}_z	\nonumber
\end{eqnarray}
From the results on purification of these states \cite{AK2}, 2D and 3D cluster states have exactly the same persistence of purifiable entanglement, but in the present case, being certain of having the optimal basis $\hat U$ is much harder. 
To compute the expectation of the witness, we consider three cases enumerated by $r$ ($r=0,1,2$). Each case corresponds to the different length of the chain $N_r=3m+2r$ ($m$ is an integer). In each of these cases, the minimal product of stabilizers to give an $N$-point correlation function is $(m+r)$ \footnote{$r=1$ must be treated as a special case since there are two such products. However, this only makes a difference to the $m=0$ case.}. Further $N$-body correlators can be constructed by multiplying by pairs of operators. For example, for $r=0$, the basic product is $K_2K_5K_8\ldots K_{N-1}$, and further products can be constructed by multiplying terms $K_{3s}K_{3s+1}$, or using the end terms $K_1$ and $K_N$. Thus the value of $T_{\vec{k}}$ is simply a power of $\tanh(\frac{\beta}{2})$, where the power is the number of products that have been used. As our Hamiltonian is defined, the best basis varies with position. For $r=0$, it is given by $\sigma_z$ and $\sigma_y$ for the end terms, $\sigma_x, \sigma_y$ at positions $3s+2$ and $\sigma_z,\sigma_x$ otherwise. Thus, we find that
\begin{eqnarray}
&&\sum_{\vec{k}}|T_{\vec{k}}|=\tanh^{r+m}(\beta/2)\times\nonumber\\
&&\times(1+\tanh^2(\beta/2))^{m+r-1}(1+\tanh(\beta/2))^{2-r}
\end{eqnarray}
and that $\sum T_{\vec{k}}^2$ is given by the same expression, but replacing all the $\tanh(\beta/2)$ with $\tanh^2(\beta/2)$. In the limit of large $N$, we find that the critical values of $\beta$, corresponding to $W_A$ and $W_B$ read $\beta_{cr}^B=2.35, \beta_{cr}^A= 1.67.$

Therefore, above the temperature $T_{cr}^B = (k_B \beta_{cr}^B)^{-1}$, the thermal cluster state does not violate two setting WWZB inequalities yet is entangled until at least $T^A_{cr}=(k_B\beta_{cr}^A)^{-1}$, above which, the entanglement witness $W_{\hat{U}}$ fails to detect entanglement. However, we know from Ref.~\cite{AK2} that purifiable entanglement exists below the temperature given by $\beta=\ln(\sqrt{2}+1)$. Thus, for all $N$, our bound finds a lower temperature than the critical temperature for purification, and hence, does not detect any bound entanglement, in contrast to the GHZ state.

It was experimentally shown based on the results in the Ref. \cite{ent-cluster} that cluster states violate some Bell inequalities \cite{Zeilinger}. Interestingly, these Bell inequalities are not violated by the GHZ state, which explains why there is a range of temperatures for which the thermal cluster state does not violate two-setting WWZB inequalities.    

\subsection{Thermal Bose-Hubbard with Single Excitation}

Moving away from stabilizer Hamiltonians, consider a $1$-dimensional regular array of $N$ lattice sites with periodic boundary conditions in which we place a single particle (the extension to $d$-dimensional lattices is trivial because we are only using a single excitation). This particle is free to hop between nearest-neighbour sites with a constant hopping amplitude. We can define a basis $\ket{n}$, denoting that the particle is on the $n^{th}$ qubit. 

The eigenstates of such a system are readily expressed as
$$
\ket{\psi_m}=\frac{1}{\sqrt{N}}\sum_{n=1}^Ne^{2\pi imn/N}\ket{n},
$$
and the eigenvalues are 
$
E_{m}=2\cos\left(\frac{2\pi m}{N}\right).
$

The thermal state of the system in this ``position" representation therefore reads
\begin{equation}
\rho=\frac{\sum_{r,s,m=1}^N e^{-\beta E_m}\ket{r}\bra{s}e^{2\pi i(r-s)m/N}}{Z(\beta)},
\end{equation}
where $Z(\beta)=N\sum_m e^{-\beta E_m}$ is the partition function. At this stage, on each site, we associate the presence/absence of a particle with the qubit levels $\ket{1}/\ket{0}$. The interpretation of whether any observed violation of the entanglement witness is really entanglement has been discussed elsewhere \cite{single}. Instead of engaging in a complete analysis of the correlators, we can trivially observe that $T_{zz\ldots z}=-1$, and therefore $W_B>1$. This arises because we know that $\Tr(\rho)=1$, and $Z$ behaves exactly like $\identity$, except that a negative sign is introduced in the presence of an odd number of excitations i.e.~to all terms. Thus, entanglement persists at all temperatures. However, in contrast to the GHZ state, this entanglement is always purifiable, as can be proven by demonstrating an explicit purification protocol. Let us perform $Z$-measurements on all sites but a particular nearest-neighbour pair, $r$ and $r+1$, and post select on all measurement results being $\ket{0}$. This leaves the density matrix
$$
\rho_{r,r+1}=\frac{\sum_me^{-\beta E_m}\left(\begin{array}{cc}
1 & e^{2\pi im/N} \\
e^{-2\pi im/N} & 1
\end{array}\right)}{2\sum_m e^{-\beta E_m}}.
$$
The fidelity with the singlet state $(\ket{r}-\ket{r+1})/\sqrt{2}$ is
$$
F=\half-\frac{\sum_me^{-\beta E_m}E_m}{2\sum_me^{-\beta E_m}}.
$$
Note that the terms with negative energy have largest weight, and therefore $F>\half$, and so it can be purified to the perfect singlet. By symmetry this can happen between all nearest neighbour sites, which is sufficient to reproduce any desired output state.

\section{Summary}

We have investigated a family $W_{\hat U}$ of $N$-qubit entanglement witnesses which are functions of $N$-point correlators.  We have derived a lower bound $Tr(W_B\rho)$ for $\Tr(W_{\hat U}\rho)$ ($\rho$ is an arbitrary $N$-qubit state) and demonstrated that the condition $Tr(W_B\rho)=1$ separates states that do not violate two-setting WWZB-type Bell inequalities from the entangled states violating multi-setting WWZB type Bell inequalities. Both $W_B$ and the stronger witness $W_A$ detect that if $\Tr(W\rho)>2^{M+1}$, at least $M+1$ qubits are entangled. Similarly, above certain thresholds, the presence of genuine multipartite entanglement can be detected.

The family  of witnesses $W_{\hat U}$ has been tested on the thermal stabilizer states (GHZ Hamiltonian and cluster state Hamiltonian) as well as on the one excitation Bose-Hubbard thermal state. 
In the case of the thermal GHZ stabilizer state, we have found that in the thermodynamic limit one can always detect entanglement with the help of $W_{\hat U}$ at any finite temperature, although it is known that there is a finite regime of temperatures for which the state is entangled and purifiable. Hence, the state is bound entangled. Contrastingly, when we considered the thermal cluster state, we found a critical temperature $T^A_{cr}$ below which entanglement is detected. The critical temperature due to the witness $W_B$ also shows that if we're below $T^B_{cr}<T^A_{cr}$, we have entanglement. However, above $T^B_{cr}$, no two-setting WWZB inequality is ever violated. In the regime between these two temperatures, the state violates multi-setting WWZB Bell inequalities.  Finally, we have also examined the thermal one excitation Bose-Hubbard model, which always violates multi-setting WWZB Bell inequalities for any finite temperature, although we have argued that in this case, the entanglement is always purifiable.

It would be interesting to apply the entanglement witnesses $W_{\hat{U}}$ to thermal states of spin Hamiltonians of ferromagnetic and anti-ferromagnetic type as well as to other models used in condensed matter physics. Another interesting question is whether the witnesses derived in the paper can be expressed as a function of the partition function and its higher order derivatives. Higher order derivatives of the partition function have not yet found applications in condensed matter physics. 

\acknowledgments
ASK is supported by Clare College, Cambridge. DK would like to thank Valerio Scarani, Marcin Wiesniak and L. C. Kwek for interesting discussions. This work is supported by the National Research Foundation \& Ministry of Education, Singapore.

\appendix

\section{Derivation of $W_B$}

In this appendix, we must prove that for our choice of $b_{\vec k}$ in Eqn.~(\ref{eqn:W_b}), we get the relation in Eqn.~(\ref{eqn:W_b2}). It is convenient to write $W$ in the Pauli basis, i.e., the basis consisting of elements $\sigma_{l_1}\otimes\dots\otimes\sigma_{l_N}$ with $\sigma_0 =1,\sigma_1=\sigma_x,\sigma_2=\sigma_y,\sigma_3=\sigma_z$. The coefficients of the expansion $\mu_{l_1\dots l_N}$ in this basis read
\begin{eqnarray}
&&\mu_{l_1\dots l_N}=\Tr(\sigma_{l_1}\otimes\dots\otimes\sigma_{l_N}W)=\nonumber\\
&&  \sum_{\vec{k}}b_{\vec{k}}\text{Re}\left(\langle\vec{k}'|\sigma_{l_1}\otimes\dots\otimes \sigma_{l_N}|\vec{k}\rangle\right).
\end{eqnarray}
Note that $\text{Re}\left(\langle\vec{k}'|\sigma_{l_1}\otimes\dots\otimes \sigma_{l_N}|\vec{k}\rangle\right)$ vanishes whenever at least one of the indices $l_j=0,3$. Thus, let us define the vector $\vec{l}=(l_1\dots l_N)$ such that $l_j=0(1)$ corresponds to $\sigma_x(\sigma_y)$ at the $j$th position. Thus we find the relation
\begin{equation}
\sigma_{\vec{l}}|\vec{k}\rangle =  (-1)^{\vec{k}\cdot\vec{l}} (i)^{|\vec{l}|}|\vec{k}'\rangle
\end{equation}
and the entanglement witnesses for a particular choice of $\hat U$, $W_{\hat{U}}$ can be written as 
\begin{equation}
W_{\hat{U}} =\frac{1}{2^N} \sum_{\vec{k},\vec{l}}(-1)^{\vec{k}\cdot\vec{l}}b_{\vec{k}}\cos{\left(\frac{\pi}{2}|\vec{l}|\right)}\hat{U}\sigma_{\vec{l}}\hat{U}^{\dagger}.
\end{equation}
Consequently, the trace of $W_{\hat{U}}$ with an arbitrary density operator $\rho$ reads
\begin{equation}
\Tr(W_{\hat{U}}\rho) = \frac{1}{2^N}\sum_{\vec{k}}b_{\vec{k}}\lambda_{\vec{k}}, \label{opt}.
\end{equation}
The number $T_{\vec{l}}= \Tr(\hat{U}\sigma_{\vec{l}}\hat{U}^{\dagger}\rho)$ is the average value of spin measurements along the directions given by $O_{n}\hat{x},O_n\hat{y}$ ($n=1,\dots,N$), where $O_n$ is an orthogonal representation of $U_n$.  We see that the only relevant $N$-point correlation functions are those for which $|\vec{l}|$, i.e., the number of $\sigma_y$ in $\sigma_{\vec{l}}$ before the local unitary operation $\hat{U}$, is an even number. 

To derive $W_B$ we put 
$b_{\vec{k}} = 2^N\lambda_{\vec{k}} \left(\sum_{\vec{l}} \lambda_{\vec{l}}\right)^{-1}$, which still satisfies Eq.~(\ref{wit}), and is hence a sub-optimal choice of witness. The result is
\begin{equation}
\max_{\hat{U}}\left(\Tr(W_{\hat{U}}\rho)\right)=\max_{\hat{U}}\left(\frac{\sum_{\vec{k}}\lambda_{\vec{k}}^2}{\sum_{\vec{k}}|\lambda_{\vec{k}}|}\right).
\end{equation}
By definition $\lambda_{\vec{k}}=2^{N-1} \Tr\left(\hat{U}^{\dagger}\rho\hat{U}(Q^+_{\vec{k}}-Q^-_{\vec{k}})\right)$ thus 
\begin{equation}
\sum_{\vec{k}}|\lambda_{\vec{k}}|\leq 2^{N-1}\sum_{\vec{k}}\Tr\left(\hat{U}^{\dagger}\rho\hat{U} (Q^+_{\vec{k}}+Q^-_{\vec{k}})\right) = 2^N,
\end{equation}
where we have used the completeness of the operators $Q^{\pm}_{\vec{k}}$, except that we remember that each is counted twice; once for $k$ and once for $k'$. Moreover, 
\begin{eqnarray}
\sum_{\vec{k}}\lambda_{\vec{k}}^2 &=& \sum_{\vec{l},\vec{m},\vec{k}}(-1)^{\vec{k}\cdot(\vec{l}+\vec{m})}\cos{\left(\frac{\pi|\vec{l}|}{2}\right)}\cos{\left(\frac{\pi|\vec{m}|}{2}\right)}T_{\vec{l}}T_{\vec{m}}\nonumber\\
&=&2^N\sum_{\vec{l}}\cos^2{\left(\frac{\pi|\vec{l}|}{2}\right)}T_{\vec{l}}^2=2^N\sum_{\vec{l}\in even}T_{\vec{l}}^2,
\end{eqnarray}
where we have made use of the formula $\sum_{\vec{k}}(-1)^{\vec{k}\cdot(\vec{l}+\vec{m})}=2^N\delta_{\vec{k},\vec{m}}$. The notation $\vec{l}\in even$ means summation only over vectors $\vec{l}$ with even Hamming weight. Combining the last two equations we finally get 
\begin{equation}
W_B=\max_{\hat{U}}\left(\sum_{\vec{l}\in even}T^2_{\vec{l}}\right),
\end{equation}
as required.

\end{document}